\RequirePackage{lineno} 
\documentclass[twocolumn,aps,prl,superscriptaddress,amsfonts]{revtex4-1}

\usepackage{amsmath}
\usepackage{graphicx}
\usepackage{bm}
\usepackage{array}
\usepackage{dcolumn}
\usepackage{hepparticles}
\usepackage{heppennames}
\usepackage{hepnicenames}
\usepackage{epstopdf}
\newcommand{\up}{\ensuremath{\left|\uparrow\right\rangle}}
\newcommand{\down}{\ensuremath{\left|\downarrow\right\rangle}}

\newcommand{\Yb}{\ensuremath{^{171}\mathrm{Yb}^+}}

\newcommand{\ket}[1]{\ensuremath{\left|#1\right\rangle}}

\usepackage[usenames,dvipsnames]{xcolor}
\usepackage{soul}

\begin{document}
\def\e{\epsilon}
\def\d{\downarrow}
\def\u{\uparrow}
\def\e{\mathcal{E}}
\def\ba{\begin{eqnarray}}
\def\ea{\end{eqnarray}}
\def\beq{\begin{equation}}
\def\eeq{\end{equation}}
\title{Co-Designing a Scalable Quantum Computer with Trapped Atomic Ions}

\author{K. R. Brown}
\affiliation{Schools of Chemistry and Biochemistry, Computational Science and Engineering, and Physics, Georgia Institute of Technology, Atlanta, GA 30332}
\author{J. Kim}
\affiliation{Department of Electrical and Computer Engineering, Duke University, Durham, NC  27708}
\author{C. Monroe}
\affiliation{Joint Quantum Institute, Joint Center for Quantum Information and Computer Science, and Department of Physics, University of Maryland, College Park, MD 20742}

\date{\today}

\begin{abstract}
The first generation of quantum computers are on the horizon, fabricated from quantum hardware platforms that may soon be able to tackle certain tasks that cannot be performed or modelled with conventional computers.  These quantum devices will not likely be universal or fully programmable, but special-purpose processors whose hardware will be tightly co-designed with particular target applications.  Trapped atomic ions are a leading platform for first generation quantum computers, but are also fundamentally scalable to more powerful general purpose devices in future generations.  This is because trapped ion qubits are atomic clock standards that can be made identical to a part in $\rm{10^{15}}$, and their quantum circuit connectivity can be reconfigured through the use of external fields, without modifying the arrangement or architecture of the qubits themselves.  In this article we show how a modular quantum computer of any size can be engineered from ion crystals, and how the wiring between ion trap qubits can be tailored to a variety of applications and quantum computing protocols.
\end{abstract}

\maketitle

Quantum information processors have the potential to perform computational tasks that are difficult or impossible using conventional modes of computing \cite{Feynman82,MikeAndIke, QC}.  In a radical departure from classical information, the qubits of a quantum computer can simultaneously store bit values 0 and 1, and when measured they probabilistically assume definite states. Many interacting qubits, isolated from their environment, can represent huge amounts of information: there are exponentially many binary numbers that can co-exist, with entangled qubit correlations that can behave as invisible wires between the qubits.  Even in the face of Moore's Law, or the doubling in conventional computer power every year or two, the complexity of massively entangled quantum states of just a few hundred qubits can easily eclipse the capacity of classical information processing \cite{VanMeter2006}.  There are but a few known applications that exploit this quantum advantage, such as Shor's factoring algorithm \cite{Shor1994}, and future quantum information processors will likely be applied to special purpose applications.  On the other hand, a quantum computer has not yet been built, so new quantum applications and algorithms will likely follow from the evolution and capability of quantum hardware.

In the 20 years since the advent of Shor's algorithm \cite{Shor1994} and the discovery of quantum error correction \cite{Shor1995EC,Calderbank96,Steane1996}, there has been remarkable progress in demonstrations of entangling quantum gates on less than 10 qubits in certain physical systems.  Current efforts aim to scale to hundreds, thousands or even millions of interacting qubits. Unlike the classical scaling of bits and logic gates however, large quantum systems are not comparable to the behavior of just a few qubits. Just because 2 or 4 qubits can be completely controlled with negligible errors does not mean that this system can readily scale to $>100$ qubits.

In the last few years, two particular quantum hardware platforms have emerged as the leading candidates for scaling to intereting numbers of qubits: trapped atomic ions \cite{WinelandBlatt08, MonroeKimScience} and superconducting circuits \cite{JJReview, MartinisPRL, IBM}.  These technologies will likely both be built out in coming years, and may find complementary uses.  Superconducing circuitry exploits the significant advantages of modern lithography and fabrication technologies: it can be integrated on a solid-state platform and many qubits can simply be printed on a chip.  However, they suffer from inhomogeneities and decoherence, as no two superconducting qubits are the same, and their connectivity cannot be reconfigured without replacing the chip or modifying the wires connecting them within a very low temperature environment.  Trapped atomic ions, on the other hand, feature virtually identical qubits, and their wiring can be reconfigured by modifying externally applied electromagnetic fields.  However, atomic qubit switching speeds are generally much slower than solid state devices, and the development of engineering infrastructure for trapped ion quantum computers and the mitigation of noise and decoherence from the applied control fields is just beginning.

In this paper, we anticipate the upcoming engineering efforts on trapped atomic ions for quantum computing, and highlight their reconfigurable quantum circuit connectivity as a flexible platform to be applied to a wide range of potential quantum applications.  This path to scaling to thousands or more qubits will almost certainly involve the concept of architectual co-design \cite{codesign}, where algorithms and applications are invented alongside the development of trapped ion hardware, and the laboratory engineers fabricate an ion trap architecture that is well-adapted to certain types of quantum circuit applications.

\section*{Ion Trap Qubits and Wires}
\begin{figure}
\includegraphics[width=1.0\linewidth]{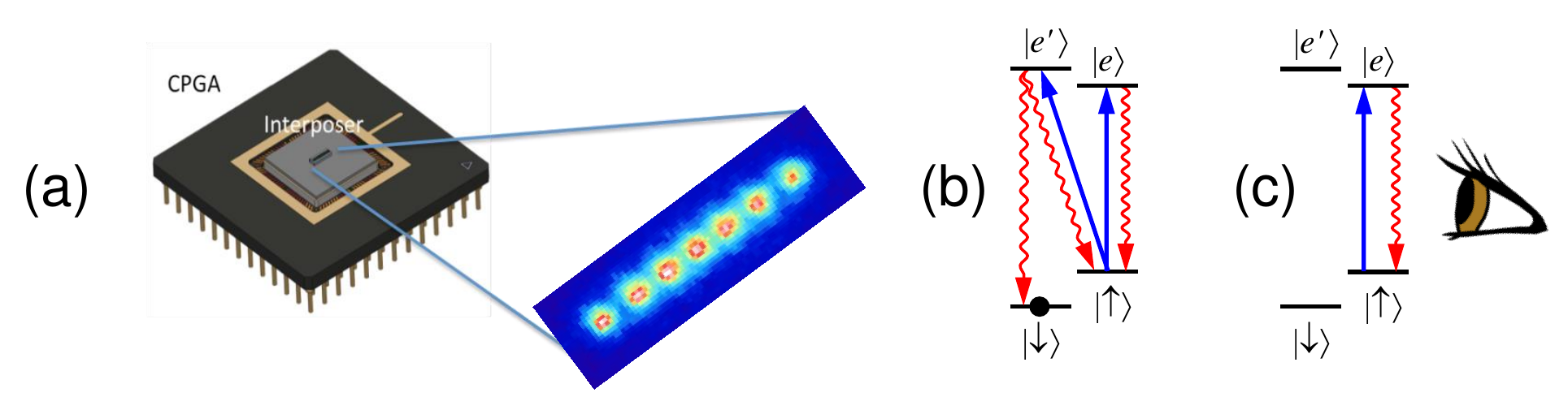}
\caption{(a) Schematic of silicon chip trap mounted on a ceramic pin grid array (CPGA) carrier with raised interposer,
confining atomic ions that hover $\sim75$ $\mu$m above the surface. The inset is an image of 7 atomic ytterbium (\Yb)  ions arranged in a linear crystal and laser-cooled to be nearly at rest. The few-micrometer separation between ions is determined by a balance between the external confinement force and Coulomb repulsion. 
(b-c) Reduced energy level diagram of a single \Yb atomic ion, showing the atomic hyperfine levels \up \, and \down \, that represent a qubit.  The electronic excited states $\ket{e}$ and $\ket{e'}$ are separated from the ground states by an energy corresponding to an optical wavelength of $369.53$ nm, and applied laser radiation (blue arrows) drives these transitions for (b) initialization to state \down, and (c) fluorescence detection of  the qubit state (\up = fluorescence, \down = no fluorescence).}
\label{ions}
\end{figure}

Atomic ions can be confined in free space with electromagnetic fields supplied by nearby electrodes.  The linear radiofrequency (rf) trap is the typical choice for quantum information applications \cite{WinelandBlatt08, MonroeKimScience}.  When the ions are laser-cooled to bottom of the trapping potential, they form a linear crystal of qubits, with the Coulomb repulsion balancing the external confinement force, as shown in Fig. \ref{ions}a. Ions are typically loaded into traps by generating neutral atoms of the desired element and ionizing the atoms once in the trapping volume.  Ion trap depths are usually much larger than room  temperature, so rare collisions  with background gas do not necessarily eject the ion from the trap, but they can temporarily break up the crystal and scramble the qubits.  Under typical ultra-high-vacuum conditions, these qubit interruptions occur roughly once per hour per ion, but cryogenic vacuum chambers can reduce the collision rate by orders of magnitude, where qubits may last months or longer.

Qubits stored in trapped atomic ions are represented by two stable electronic levels within each ion, often represented as an effective spin with the two states \down \,  and \up \, corresponding to bit values $0$ and $1$.  The qubits can be initialized and detected with nearly perfect accuracy using conventional optical pumping and state-dependent fluorescence techniques \cite{DidiRMP}. This restricts the atomic species of trapped ion qubits to those with simple electronic structure (e.g., those with a single valence electron: Be$^+$, Mg$^+$, Ca$^+$, Sr$^+$, Ba$^+$, Zn$^+$, Hg$^+$, Cd$^+$, and Yb$^+$).  

Figs. \ref{ions}b and c show the reduced energy level diagram of \Yb, where the qubit levels \down \, and \up \, are represented by the stable hyperfine levels of electron/nuclear spin in the ground electronic state, separated by frequency $\nu_{HF} =12.642 812$ GHz.  Such states form an excellent freqency standard, and coherence times $>1000$ s have been observed \cite{Bollinger1991,Fisk1997}.  The optically-excited electronic states \ket{e} and \ket{e'} are themselves split by a hyperfine coupling and separated from the ground states by an optical interval.  Laser radiation tuned just below resonance in these optical transitions allows Doppler laser cooling to confine ions near the bottom of the trap.  Other more sophisticated forms of laser cooling can bring the ions to nearly at rest in the trap 
\cite{DidiRMP}. When laser beams resonant with both $\up \leftrightarrow \ket{e}$ and $\up \leftrightarrow \ket{e'}$ transitions are applied, the ion rapidly falls into the state \down \, and no longer interacts with the light field (Fig. \ref{ions}b).  This optical pumping technique allows the initialization of a qubit with essentially $100\%$ fidelity. When a single laser resonant with the transition $\up \leftrightarrow \ket{e}$ is applied, the closed cycling optical transition causes an ion in the \up \,  state to fluoresce strongly at a rate scaled by the excited state radiative linewidth $\gamma \sim 2\pi \times 10$ MHz, while an ion in the \down \,  state stays dark, because the laser is far from its resonance (Fig. \ref{ions}c).  The collection of even a small fraction of this fluorecence thus allows for the detection of the atomic qubit state with near-perfect efficiency, with integration times as low as $\sim 20$ $\mu$s \cite{Noek2013}.  Other atomic species have similar initialization/detection schemes.

\subsection*{Wiring atomic qubits with the Coulomb interaction}
\begin{figure}
\includegraphics[width=1.0\linewidth]{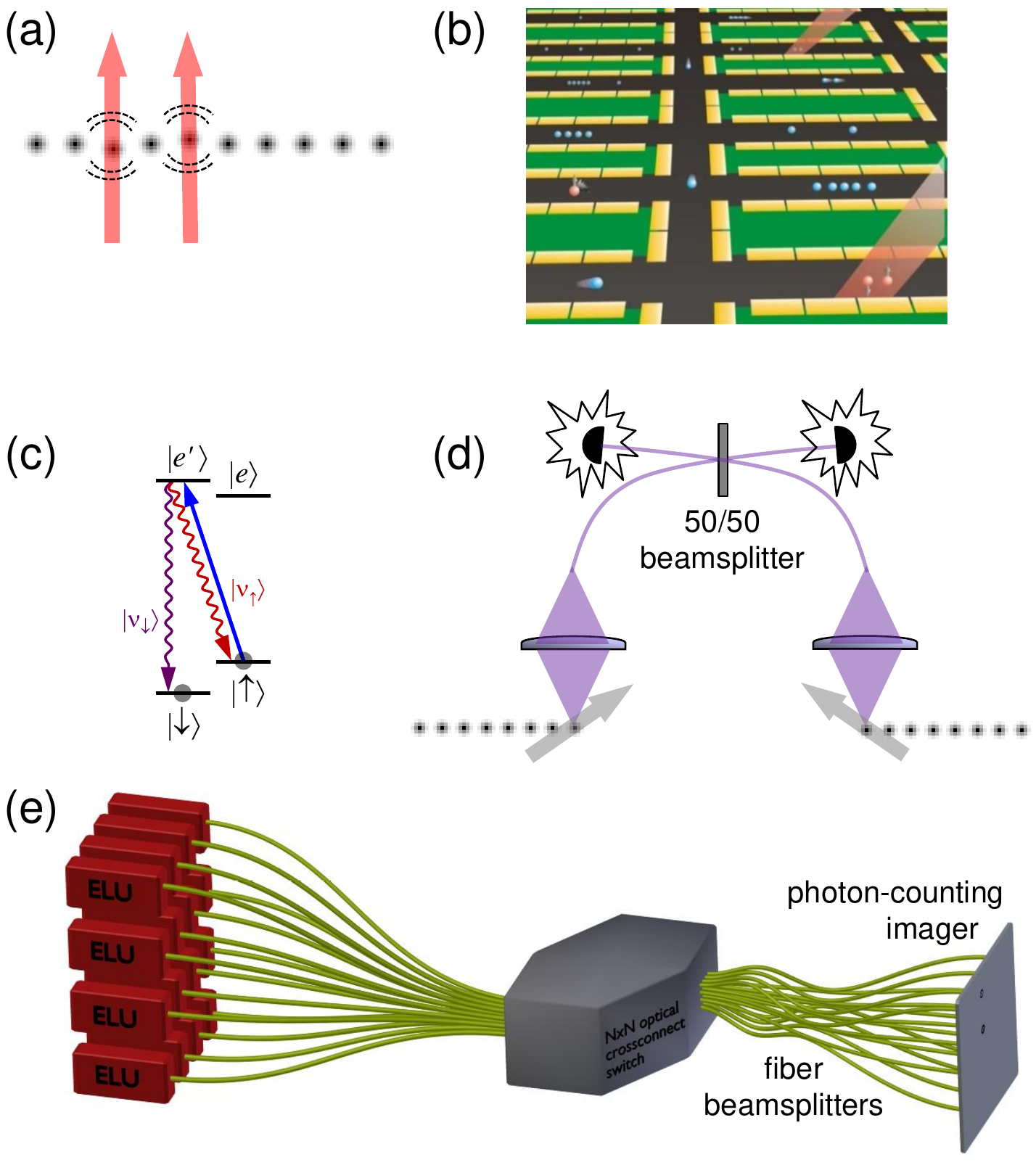}
\caption{(a) Qubit state-dependent forces on individual ions in a chain provide a coupling mechanism between qubits for the operation of entangling quantum logic gates.  Optical dipole forces (indicated in red) displace two ions depending upon their qubit states, and the resulting modulation of the Coulomb interaction allows the implementation of entangling quantum gates between these two ions.  This type of link can be generalized to operations involving any number of qubits or even the entire chain for applications in quantum simulations.
(b) Concept of a QCCD ion trap \cite{QCCD}, where ions can be shuttled between various zones for multiplexing (courtesy, NIST).
(c) Energy levels of trapped ion excited with a fast laser pulse (blue upward arrow) that produces single photon whose color, 
represented by the state $\ket{\nu_\uparrow}$ or $\ket{\nu_\downarrow}$, is entangled with the resultant qubit state \up \, or \down, respectively.  Alternative configurations allow the encoding of polarization or time-bin photonic qubits.
(d) Two ``communication" ions, immersed in a separate crystals of other ions, each produce single photons when driven by laser pulses (blue) as in (a). With some probability, the photons arrive at the $50/50$ beamsplitter and then interfere.  This ``Bell-state detection" of the photons heralds the entanglement of the trapped ion qubits.
(e) Modular distributed quantum computer. Several ELU modules, each with full control of 50-100 trapped ions through Coulomb gates, are connected through a photonic network utilizing an optical crossconnect switch, inline fiber beamsplitters and a photon-counting imager \cite{MUSIQC}. }
\label{Links}
\end{figure}

The motion of many trapped ions is coupled through their mutual Coulomb repulsion, so the qubits can therefore be linked by relating the internal qubit states to the external motion of the ions, as depicted in Fig. \ref{Links}a.  This is typically accomplished by applying qubit state-dependent optical or microwave dipole forces to the ion(s) \cite{DidiRMP,WinelandBlatt08}.  To see how this type of motional data bus works, we assume that a given ion responds to an external field $\mathcal{E}$ by experiencing an equal and opposite energy shift $\Delta E=\pm\mu\mathcal{E}$ that depends upon the qubit state through selection rules. When the field $\mathcal{E}(x)$ varies with position $x$, there is a qubit-state-dependent force of the form $F_x = \mu \mathcal{E}'(x) \hat{\sigma}_z$,  where $\mathcal{E}'(x)$ is the field gradient and $\hat{\sigma}_z$ is the Pauli matrix corresponding to the angular momentum of the qubit effective spin.  Here we neglect higher order field gradients, which is justified when the ion is laser-cooled to much less than the characteristic length scale (or wavelength) of the applied field.  For plane wave radiation coupled with wavevector $k$ and amplitude $\mathcal{E}_0$, $F_x = \hbar k\Omega\hat{\sigma}_z$, where $\hbar$ is Planck's constant and the Rabi frequency $\Omega=\mu\mathcal{E}_0/\hbar$ parametrizes the field-qubit coupling.  (For two-photon optical Raman couplings, the effective wavevector $k$ is given by the wavevector difference between the two beams \cite{DidiRMP}.)

 Because this force acts differently on the two qubit states, it couples the qubit state to the collective motion of $N$ ions, with  characteristic speed $R_{\text{gate}} = \Omega   \sqrt{\omega_{R}/\omega}$ where $\omega_R = \hbar k^2/(2Nm)$ is the recoil frequency of the ion crystal associated with field momentum $\hbar k$, $Nm$ is the total mass of the ions, and $\omega$ the frequency of harmonic oscillation of collective motion along the $x$ direction.  When this mapping affects multiple ions, entangling gates can be operated between separated ions, mediated through the motion.  There are many protocols for the creation of controlled-NOT and other gates using this coupling to the collective motion of the ions \cite{WinelandBlatt08}.  Current experiments with a few ions have realized entangled state fidelities of greater than $99.9\%$ \cite{HartyPRL2014} and operate in the range $R_{\text{gate}}/2\pi \sim 10-100$ kHz, although with ultrafast high-intensity optical fields it may be possible to operate gates in the GHz range \cite{FastGate}. 

As the number of ions $N$ in the crystal grows, the gate speed slows down as $R_{\text{gate}} \sim 1/\sqrt{N}$.  For large crystals, there will also be crosstalk between the many modes of collective motion.  Background errors such as the decoherence (heating) of the motional modes \cite{Turchette} or fluctuating fields that add random phases to the qubits will become important at longer times, thus there will be practical limits on the size of a single crystal for the performance of faithful quantum gates.   Through the use of individual optical addressing of ions \cite{Haffner08, Choi2014} and pulse-shaping techniques \cite{PulseShaping}, these errors should not be debilitating for the full control of single crytals ranging from $N=10-100$ qubits.

In order to scale beyond $\sim 50$ trapped ion qubits, we can shuttle trapped ions through space in order to couple spatially separated chains of ions in a multiplexed architecture called the quantum charge-coupled device (QCCD) \cite{QCCD} and depicted in Fig. \ref{Links}b. The QCCD architecture requires exquisite control of the atomic ion positions during shuttling and may require additional atomic ion species to act as ``refrigerator" ions to quench the excess motion from shuttling operations \cite{barrett03}.  Rudimentary versions of the QCCD idea have been employed in many quantum information applications such as teleportation and small quantum algorithms \cite{WinelandBlatt08}, and recent experiments have shown the reliable, repeatable, and coherent shuttling of ion qubits over millimeter distances in microsecond timescales \cite{Walther2012, Bowler2012} and through complex two-dimensional junctions \cite{Blakestad,Sandia,Wright13,Shu14}. The QCCD approach will help usher the development of small trapped ion quantum computers with perhaps $50-1000$ qubits.  However, scaling to many thousands or more qubits in the QCCD may be challenging due to the complexity of interconnects, diffraction of optical beams, and the extensive hardware required for qubit control.

\subsection*{Wiring atomic qubits with photons}
To scale beyond the QCCD in a modular architecture, it will likely become necessary to link separate registers of trapped ion chains with photonic interfaces \cite{DeVoe1998,Steane2000}.  This allows quantum gates to be performed between any qubits in the processor, regardless of their relative location \cite{DuanRMP,MUSIQC}, while supporting fault-tolerant error correction even in the face of photonic interconnects that succeed with small probability per attempt \cite{DuanRMP,Benjamin2013,MUSIQC}. 

A pair of trapped ion qubit modules (elementary logic units or ELUs) can be entangled with each other using propagating photons emitted by a subset of ions from each register, designated to be ``communication qubits.''  As shown in Fig. \ref{Links}c, the communication qubit is driven to an excited state with fast laser pulses so that at most one photon emerges from each qubit following appropriate radiative selection rules. When photons from two separate communication qubits are collected, mode-matched and interfered on a $50/50$ beamsplitter (Fig. \ref{Links}d), detectors on the output modes of the beamsplitter herald a Bell-state of the photons and thus the creation of entanglement between the memory qubits through entanglement-swapping \cite{Simon2003, Moehring07}.  The mean connection rate of this photonic interface is $R(F\eta_D)^2/2$, where $F$ is the fraction of light collection from each ion emitter and $\eta_D$ is the single photon detector efficiency \cite{singlephoton}.  The repetition rate $R$ of the initialization/excitation process is limited by the emission rate $\gamma$.  For typical atomic transitions into free space with $\gamma/2\pi \sim 10$ MHz, light collection fraction $F \sim 1-10\%$, and detector efficiency $\eta_D \sim 20\%$, we find typical connection rates of $\sim 100$ Hz \cite{Hucul}, but this could be dramatically improved with integrated photonics, as discussed below.  While this photonic entanglement source is probabilistic, the detected photons announce when it does succeed, and thus the heralded entanglement of the trapped ions can be subsequently used for deterministic quantum information applications \cite{KLM}.  Moreover, by performing such operations in parallel on many pairs, a FIFO (first in, first out) buffer can provide a synchronous stream of entangled pairs between the trapped ion modules that can be used as needed, thus eliminating the probabilistic nature of the connection. 

In practice, the communication qubit must be well-isolated from the memory qubits so that scattered light from the excitation laser as well as the emitted photons themselves do not disturb the spectator qubit memories.  It may be necessary to physically separate (shuttle) the communication qubit away from the others, invoking techniques from the QCCD approach, but ultimately using two different atomic species can eliminate this crosstalk \cite{Schmidt05,TanArXiv2015}, such as \Yb for memory qubits and $\ensuremath{^{138}\mathrm{Ba}^+}$ for communication qubits.  Here, the communication qubits are connected through the photonic channel, and then mapped to neighboring memory qubits though Coulomb gates as described above.

In Fig. \ref{Links}e, we show a concept of a large-scale modular trapped ion quantum computing architecture, involving individual ELU modules that host Coulomb-based quantum links within the module and can be wired to other ELU modules through photonic connections as described above.  By using a non-blocking optical crossconnect switch, the connectivity between the entire sample of qubits can be extended in order to scale up to very large numbers of ELU modules, potentially to thousands or millions of qubits.

\section*{Integration Technologies for Trapped Ion Quantum Computers}
Unlike classical solid state circuits where large-scale integration of complex information processors is readily available, practical implementation of a trapped-ion quantum processor will require development of new integration technologies and system engineering approaches. In this section, we will describe the current efforts towards such technology development.

\subsection*{Chip Traps and Optical Control of Qubits}
Reliable and reproducible fabrication of many identical ELU modules starts with the ion trap itself.  Ion trap electrode structures can be fabricated by lithographically etching semiconductor platforms such as Si/SiO$_2$ wafers and metallizing the electrodes, with positions defined to sub-micrometer precision.  The electrodes must hold high static and rf electrical potentials, with excellent insulating barriers between the electrodes, all in an ultra-high vacuum (UHV) environment.  The trap structure must also be optically open, and allow high-power laser beams to cross near the electrode surfaces to affect Coulomb gates or photonic couplings, without causing excessive light scattering.  Finally, the ions must be spatially separated from other qubit memories during initialization and measurement processes, requiring precise control over the electrical potentials over space and time in order to  shuttle ions throughout the trapping regions. 

\begin{figure}
\includegraphics[width=1.0\linewidth]{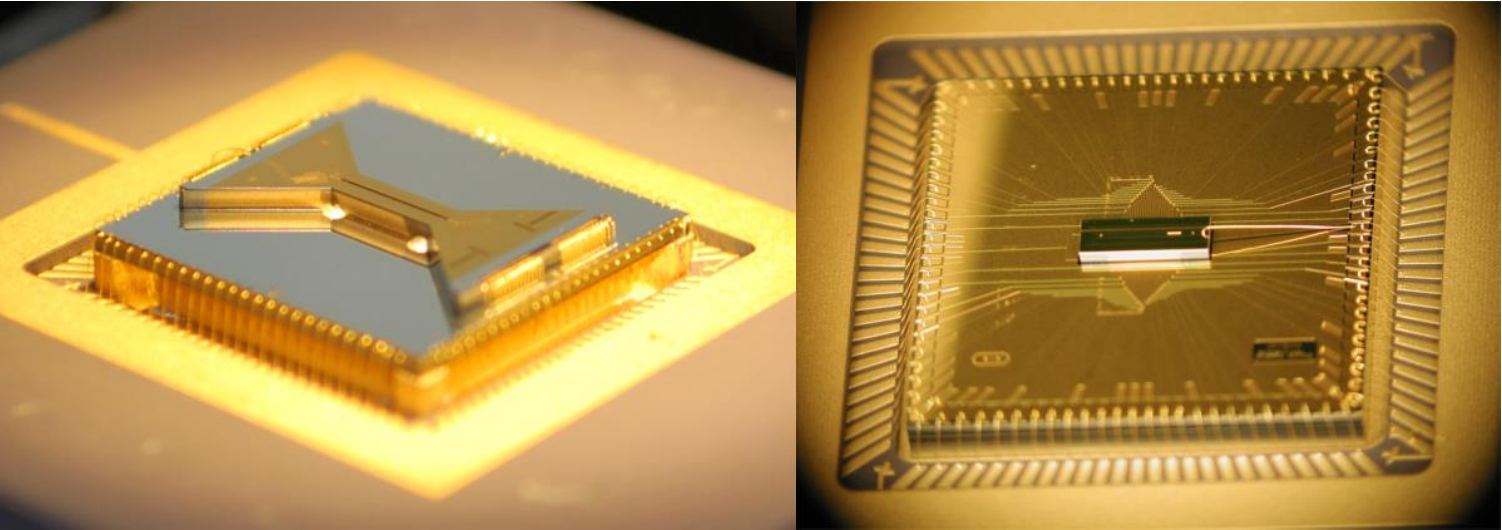}
\caption{Advanced microfabricated ion traps.  LEFT: High-optical access (HOA) trap from Sandia National Laboratories (Image courtesy of Duke University). RIGHT: Ball-grid array (BGA) trap from GTRI/Honeywell (Image courtesy of Honeywell, Inc.).}
\label{chiptrap}
\end{figure}

An approach to create the necessary electromagnetic potential to trap an ion above the surface of a chip was first suggested in 2004 \cite{NISTsurface05,KimQIC2005}, and the first monolithic semiconductor ion trap was demonstrated in 2006 \cite{Stick06}.  The design and fabrication of complex surface traps using silicon microfabrication processes has now matured, with examples of the Sandia high-optical access (HOA) trap and the GTRI/Honeywell ball-grid array (BGA) trap shown in Fig. \ref{chiptrap}.  Recent experiments have demonstrated high performance qubit measurement \cite{Noek2013} and quantum gates \cite{WangSurfGate10,MountArXiv2015,HartyPRL2014,TanArXiv2015,BallanceArXiv2015} in such microfabricated surface traps that outperform conventional manually-assembled macroscopic traps. The ability to design and simulate the electromagnetic trapping parameters prior to fabrication provides an attractive path to developing  complex trap structures that are both repeatable and produced with high yield.

Once atomic ion qubits are produced and confined to standard semiconductor trap structures, interactions between arbitrary subsets of the qubits in a single ELU or reduced sets of ions between ELUs must be gated and controlled in order to perform the desired calculations, simulations, or quantum circuit.  For local Coulomb gates as described above, we require a fixed-frequency off-resonant laser to provide dipole forces and a laser beam distribution technology that drives the desired interactions to operate a programmable and reconfigurable  quantum computer.  In the \Yb system, lasers at 355 nm are ideal for gate operations, with reliability owing to their widespread use in conventional UV lithography. For optical beam delivery, recent progress in micromirror technology \cite{Crain2014} and multi-channel acousto-optic modulator (AOM) technology developed for the optical communication and semiconductor fabrication industries \cite{32AOM} are attractive solutions, and in the coming years these devices will be tightly integrated with trapped ion systems.

\subsection*{Compact Lasers and Vacuum System Technology}
A large scale ion trap quantum computer requires several tunable laser systems to match electronic resonances in the atomic ion, with optical frequencies stable and accurate to better than $10$ kHz, a fractional precision of $\sim 10^{-10}$. Traditionally, a significant effort is dedicated to laser stabilization, with individual optical components on an optical table utilizing long optical path lengths.  Such a large footprint invariably drifts due to environmental changes (temperature, humidity, air pressure, etc.), and requires constant adjustments to keep the system operational.  Compact and stable tunable semiconductor lasers have recently been developed that provide the narrow linewidths necessary for initializing and reading out trapped ion qubits \cite{Ball2013,MountArXiv2015b,Schafer2015}. The complete optical system including the frequency stabilization can be assembled on a compact optical breadboard or a microfabricated optical bench \cite{Gates1996}. Following the footsteps of laser integration in modern dense wavelength-division multiplexed (DWDM) optical communication systems \cite{OptNetBook}, it is feasible to design and assemble a stabilized laser system where all lasers necessary for running an ion trapping experiment are packaged in a compact box that fits on an instrument rack, with fiber optic delivery to the ion trap chip.

\begin{figure}
\includegraphics[width=1.0\linewidth]{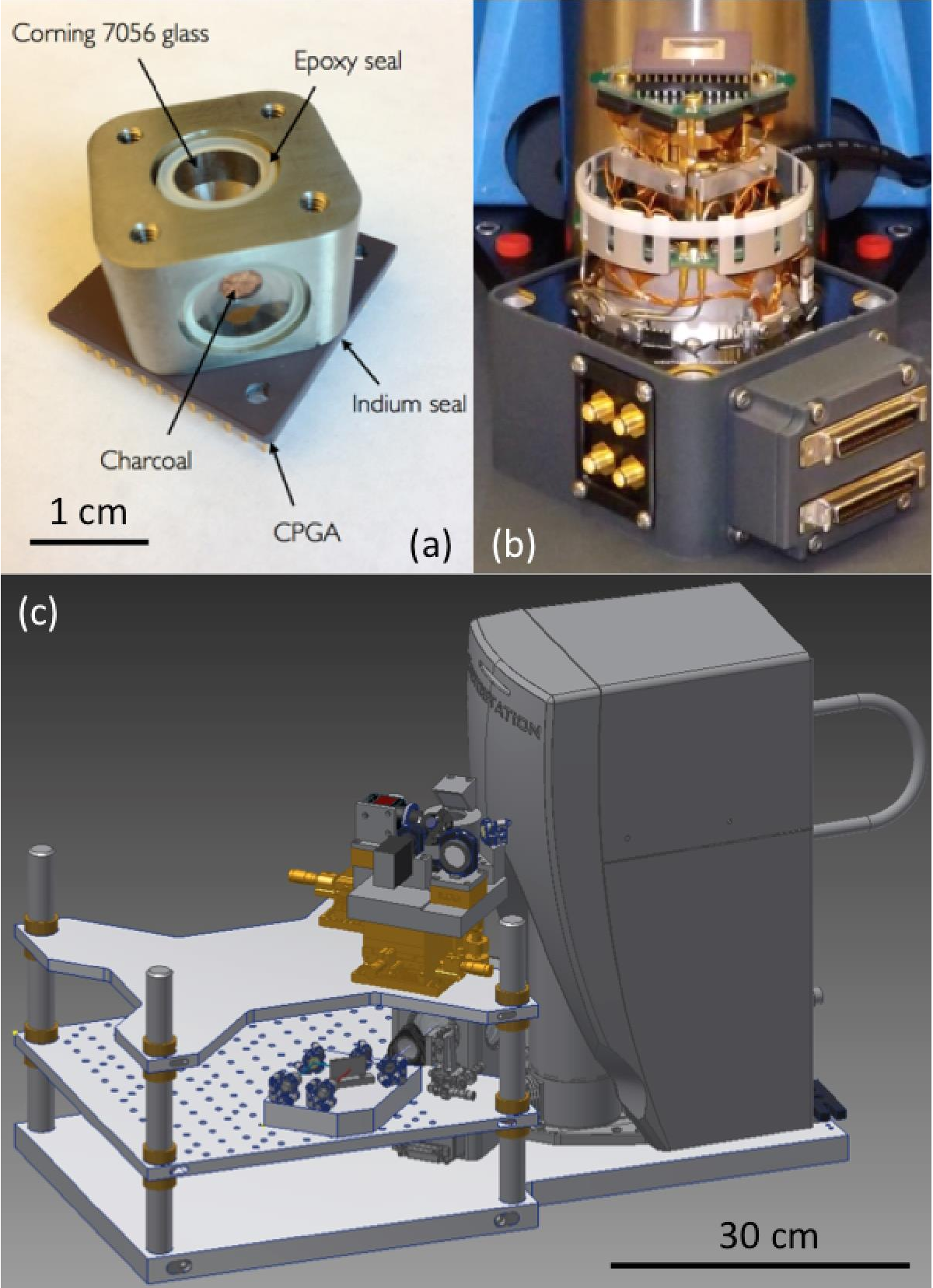}
\caption{Compact cryogenic UHV enclosure for trapped ions. (a) On-package vacuum enclosure, sealed in a UHV environment, that contains the ion trap, getter pumps, and the atomic source. (b) Upon installation and cooling in a compact cryostat, the UHV environment is established. (c) The optical components can be arranged in a compact volume around the cryostat to support the ion trap operation.}
\label{vacuum}
\end{figure}

Trapped atomic ions are suspended in an ultra-high vacuum (UHV) environment, where collisions with background molecules should be minimized for sustained operation. While careful assembly of clean UHV chamber can help create such an environment, ultimate vacuum environment might require low temperature operation ($<10$ K) \cite{Lab08a}. Closed-cycle cryogenic technology will dramatically reduce the volume and operational burden of a UHV environment, while improving the vacuum conditions for operating the trapped ion quantum computer. Figure \ref{vacuum} shows an example of a compact vacuum environment created on a ceramic package that holds the surface trap. After the trap is die-attached and wirebonded to the ceramic package, a sealed cover is assembled in a UHV environment (Fig. \ref{vacuum}a,b). The cover provides all optical access necessary to operate an ion trap, an ion source that utilizes laser ablation technique to load the trap, and getter material that will efficiently pump any residual gas molecules at low temperatures. This compact ceramic package can be installed in a closed-cycle cryostat and cooled down to cryogenic temperatures ($\sim$5K) to provide the operating environment for the ion trap quantum processor as shown in  Fig. \ref{vacuum}c. This approach is compatible with recent development in surface treatment techniques for ion traps that is shown to substantially reduce anomalous heating, which may be necessary for high-fidelity operation of multi-qubit gates \cite{HitePRL2012,Haffner2014}.

\subsection*{Photonic Technology}
Integrated optical technology will be critical for a large-scale trapped ion quantum computer.  While efficient light collection and detector arrays will be necessary for the measurement of many trapped ion qubits through state-dependent fluorescence, it will be crucial for the single-photon linking of ELU modules as discussed above.  With high numerical-aperture collection optics, $\sim10\%$ of  the emitted photons can be collected \cite{Hucul}, and $\sim50\%$ could be extracted through an optical cavity integrated with the ion trap~\cite{KimMaunzKimPRA2011}.  Highly efficient photonic Bell-state detectors with near-ideal mode-matching can be realized in fiber or waveguide beamsplitters~\cite{JenkinsOL1992} and near-unit efficiency photon detectors \cite{detectors,MarsiliNaturePhoton2013}.  Taken together, these advances may allow the linking two ELUs to approach the speed of local Coulomb-based gates ($\sim 10$  kHz).  

For large numbers of optically-networked trapped ions with many optical communication qubits, multiplexed photonic circuit elements will be necessary.  Non-blocking and transparent optical cross connect (OXC) switches with many input/output ports, developed for conventional optical communication networks and data centers \cite{Kim03,Calient03}, are well-suited for this task. Transparent optical switches establish an optical path between select input and output ports by using passive optical elements such as tilting mirrors \cite{Aksyuk03} and can guide single photons that are entangled with the trapped ion qubits to form quantum links.  These devices can also be reconfigured in real time to make parallel connections between multiple ELUs.  

The extension of the above integrated photonics technology, including detectors and waveduides to OXC switches, to the visible and blue portions of the spectrum where atomic ions respond, will be highly valuable to the trapped ion quantum optical network.  Alternatively, noiseless photonic conversion technology from visible/blue to infrared and telecom bands will play an important role, especially for long-distance quantum communication network applications.

\subsection*{Hardware and Software for Scalable Controllers}
While these integration technologies play a crucial role in developing compact, stand-alone ion trap quantum hardware, a scalable controller system is needed to run such a system. The controller system consists of hardware needed to (1) maintain the operation of frequency stabilized laser systems, (2) manage the ion position by control of rf and static voltages, (3) measure and process emitted photons with photon detectors and associated readout circuits, and (4) apply laser pulses that are generated by a digital system to prepare, measure, and manipulate qubits.  These controller systems must be designed to precisely track the amplitude and phase of the qubits used in the information processing task. It must be accompanied by control software that the user can program to instruct the quantum hardware to carry out the desired task. Both the hardware and the software for the controller should be designed for modularity and expandability, consistent with a fully coherent control of all qubits in the system. The practical scalability of the ion trap processor may ultimately be limited by the scalability of the interface between such a classical controller and the ion qubits, or how many ion qubits the controller can manage. A careful design of such controller system amounts to the ``operating system'' for the ion trap quantum processor.

\section{Applications and Opportunities for the Trapped Ion Quantum Processor}
Quantum algorithms, applications, and error correcting codes are usually designed independently of the underlying hardware, and therefore do not respect the underlying geometry of the physical system. The modular ion trap architecure has a flexible and reconfigurable connectivity that allows for the realization of arbitrary geometries with a minimal number of swaps and/or teleporation steps. In this section we examine the opportunities that such a hardware affords, beyond conventional gate-model quantum applications such as Shor's factoring algorithm.

\subsection*{Topology of Interactions}
While the spatial geometry of a crystal of laser-cooled trapped ions is typically one-dimensional, the interaction graph between qubits within a single ELU module can be fully connected and have high or undefined dimensionality. Owing to the strong long-range Coulomb interaction between the ions, quantum gates can be realized directly between distant pairs of qubits in the chain. Furthermore, multiple two-qubit gates can be performed in parallel on the chain, and multiqubit or even global entangling gates can even be performed by carefully controlling the intensity and spectrum of the lasers on all ions \cite{KorenblitNJP2012}.  Such multiqubit operations are useful for the generation of certain entangled states \cite{MS, WinelandBlatt08} and the simulation of long-range global Ising interactions between the spins \cite{Porras04, FermiSchool2014}.  In the context of such multiqubit operations, additional single qubit gates can be used to remove links in the graph \cite{Hayes14} and generate arbitrary circuits. 

\begin{figure}
\includegraphics[width=1.0\linewidth]{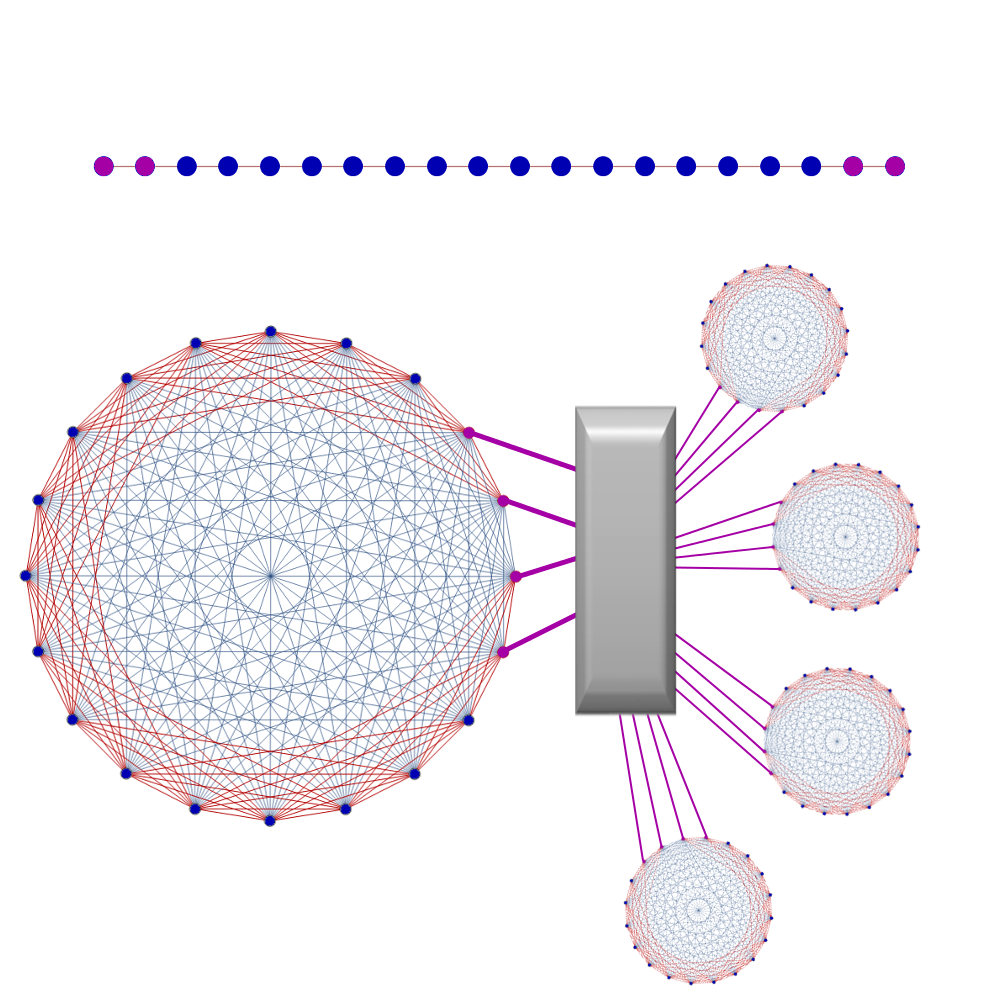}
\caption{Each ELU holds a chain of ions. The ions for remote entanglement (purple circles) are located at the ends of the chains. The interaction of the ions via the normal modes of the crystal forms a complete graph as shown by the blue lines when the ions are arranged in a circle. Fast two qubit gates based on ion proximity are labeled with red lines. The physical geometry can be observed in the break of the red connections between the remote entangling ions at the end of the chain.  Entangled pair generation between ELUs requires photonic connections (purple lines) between the ELUs and the optical networking switchyard (gray box).  The remote entanglement is reconfigurable and renewable throughout the computation. The picture shows chains of 20 ions, with 4 entangling ions, and a local fast gate distance of 4 ion spaces. Chains of 50 ions with a local gate distance of 9 ions should be possible.}
\label{connections}
\end{figure}

In the larger ion trap quantum computer, such highly-connected ELU modules are linked to other modules through photonic channels, as discussed above \cite{MUSIQC}.  This higher-level graph is determined by the density of photon-coupled ions (see Fig. \ref{connections}) and is itself dynamic and reconfigurable, leading to great advantages in the use of this type of hierarchical architecture for a host of quantum applications, and may even suggest algorithm structures that have not yet been discovered.

\subsection*{Quantum Simulation}
Quantum simulators exploit a standard well-controlled quantum system to emulate model Hamiltonians that cannot easily be understood or solved numerically \cite{Feynman82}.   The global entangling operations between trapped atomic ions are naturally suited to simulate hard quantum problems, some that may not necessarily correspond to a physical system (such as graphs that have nonlocal structures), and others with local interactions such as quantum magnetic Ising models and many body spin physics.  For instance, when executing state-dependent forces as discussed above, the applied field can be adjusted to simulate variable-range Ising models with interaction strength falling off with a power law $1/r^\alpha$ as the physical distance $r$ increases, where the exponent can be tuned between $\alpha=0$ (infinite-range) and $\alpha=3$ (dipole-dipole) \cite{Porras04, Kim09, Britton2012}.  Such simulations could assist our understanding of models of exotic materials (such as high-temperature superconductors), or even stimulate the search for new material properties that have not yet been observed.

Determining the equilibrium behavior (ground state) of spins subject to an arbitrary Ising coupling graph and local field terms is an NP-hard problem \cite{NPHard}. The quantum adiabatic algorithm attempts to find the ground state by starting in a strong transverse field and then adiabatically switching it off \cite{Farhi01}. Since the minimum energy gap to the excited states is not known for the most difficult problems, ``adiabatic" in this case means slow enough to consistently produce the same final state. The quantum annealing algorithm is similar in style but interaction with a thermal bath complicates the dynamics \cite{Boixo14, AlbashEPJ2015}. Although a sufficiently cold bath could improve the possibility of finding the ground state, this is not always the case. Large-scale superconducting systems fabricated by D-Wave have generated significant controversy \cite{Shin2014}, and the results to date are consistent with simulations of an open-system quantum annealer having limited use \cite{Boixo14, Pudenz2015, AlbashEPJ2015, Albash2015, Katzgraber2015, Ronnow2014}. The main limitations are limited precision and stability in the coupling parameters, two-qubit couplings of fixed type, a simple 2D static network structure, and a temperature that is much larger than the characteristic energy scales in the system. As the gaps are expected to shrink polynomially with the system size for good problems, the D-Wave system thus relies on thermal effects.  While trapped ion qubit couplings are roughly 1000 times smaller than superconducting systems in absolute terms, ion trap architectures promise significant advantages: higher coupling precision, two-qubit couplings of different forms, arbitrary and reconfigurable network structure, and an effective zero temperature environment \cite{WinelandBlatt08}.  While key challenges remain in such an ion trap quantum adiabatic processor, such as stable laser delivery and the engineering challenges of fabricating large-scale ion trap chips as discussed above, trapped atomic ions are well-suited to the generic problem of quantum adiabatic algorithms.


\subsection*{Machine Learning and the Boltzmann Machine}
Many models of artificial intelligence or machine learning are inspired by natural neural networks.  In a model where the communication between neurons is bi-directional, the problem of determining the state of output neurons can be mapped to calculating the thermal distribution of an  Ising model. The process of learning is  the strengthening and weaking of connections between neurons, so that the output neurons optimally classify the signal received by the input neurons.  In the model, learning is achieved by tuning the Ising model parameters to optimize the classification.  This model of learning, known as a ``Boltzmann machine" \cite{Bengio2013}, can be simulated using an adiabatic/annealing quantum protocol well-suited to the ion trap architecture.

The Boltzmann machine is borrowed from the field of  machine learning, uses the thermal distribution of an Ising model to make classifications \cite{Bengio2013}. These methods have had a recent renaissance due to increased computational power and large data sets making deep learning both practical and useful. The optimization procedure involves inputting training data and observing the classifier labels. The Ising couplings between the spins are adjusted until the machine generates an optimal classification. Quantum annealing or the quantum adiabatic method can be used to determine the classification of data. This may result in a different set of optimal coupling values which can also be tested on classical machines. Classical algorithms tend to use layers of spins where each layer can have arbitrary connections with the next layer: two layers is a reduced Boltzmann machine, many layers is a deep reduced Boltzmann machine. A full Boltzmann machine  allows for connections between any spin.  Quantum algorithms are predicted to speed up the tuning of the machine and have already provided insight into new classical algorithms \cite{Wiebe2014,Wiebe2015}. Both full and reduced Boltzmann machines can be implemented with our modular system.  The naturally connectivity suggest a different class of Boltzmann machines where the outputs of one full Boltzmann machine can be teleported to the inputs of a second full Boltzmann machine.

\subsection*{Quantum Error Correction}

Quantum error correction (QEC) extends known classical error correction techniques to quantum information by redundantly encoding effective logical qubits in a large number of raw physical qubits, and through measurements or other dissipative techniques, the logical qubit errors can be made smaller \cite{MikeAndIke, QC}.  When the error per physical qubit drops below a particular threshold, there are schemes for making the system fault tolerant, even for arbitrarily long computations \cite{GottesmanFTEC}.  

Calderbank-Steane-Shor (CSS) codes \cite{Shor1995EC, Steane1996, Calderbank96}, a common class of QEC, are constructed from two classical codes. The check operators of the first code measure the parities of sets of qubits in the computational basis to correct bit flips.  The second code checks the parities in the Hadamard basis to correct phase flips.  For each error type, we can create a graph based on the interaction between the data qubits and ancillary qubits that measure the check operator.  We refer to this graph as the QEC graph.   For simplicity, we assume the check is performed using a single ancilla qubit that interacts with a set of data qubits. The nodes of the graph are the data qubits and ancilla qubits.  The edges are between the data qubits and the ancilla qubits that query them. 

\begin{figure}
\includegraphics[width=1.0\linewidth]{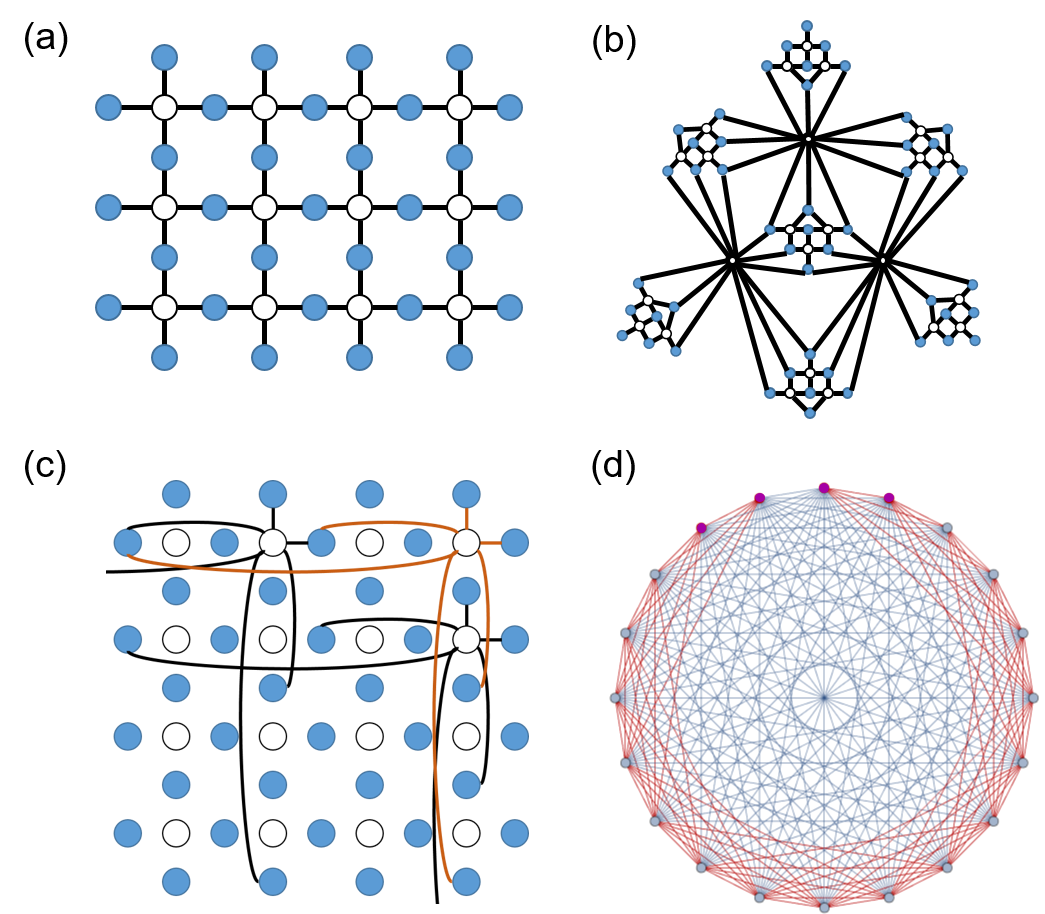}
\caption{QEC graphs for (a) a patch of the surface code, (b) a concatenated Steane [7,1,3] code, and (c) a patch of the hypergraph code. For the hypergraph code only three stabilizers are shown for clarity. The full graph can be formed by translating these operators. The blue circles indicate data qubit and the white circles indicate syndrome bits. The syndromes are described by links between the data and syndrome bits. The surface code has a natural two-dimensional layout with low-weight check operators (low degree nodes). Concatenated codes have check operators whose weight grows with the system size. Hypergaph codes have fixed weight operators but cannot be represented as local connections in two dimensions. (d) All graphs with $n$ nodes are subgraphs of the complete graph of n nodes.  The complete graph is the natural qubit connectivity of ions in an ELU as depicted here (see Fig. \ref{connections} for more details.)}
\label{QEC_Graph}
\end{figure}

Mapping a quantum error correcting code to a specific physical architecture is tantamount to mapping the QEC graph on to the graph of physical interactions. The surface code and two-dimenstional color codes are attractive  because  their QEC graph is planar, meaning that the nodes can be arranged on a surface such that no edges intersect.  It is simple to map these architectures to a two-dimensional array of qubit interactions commonly found in solid state implementations of quantum computers as shown in Fig. \ref{QEC_Graph}.

In addition to having a local QEC graph,  the surface code also has a promising error threshold of around 1\%  \cite{DKLP, Raussendorf2007, Fowler2012}. However, surface codes have two shortcomings: (i) the code is "zero rate" meaning that the number of physical qubits required per logical qubit increases without bound as the logical error is reduced, and (ii) universal computation is not possible without gate teleportation, which is accomplished by the production and consumption of magic states \cite{BravyiMagic}.  These two properties result in an increase in resource costs for large scale quantum computation relative to an ideal code.  At present, no ideal quantum code is known which has a high threshold, finite rate, and allows for magic-state free universal computation.

Codes with finite rate have must have check operators that are spatially non-local.  Codes with spatially local check operators are topological codes and  reducing the error is equivalent to increasing the physical size of the code fabric by adding qubits; this volume expansion does not change the topology, and therefore cannot increase the number of qubits.  In contrast, consider the hyper-bicycle codes as an example of a finite rate code. The check operators still act on a sparse set of data qubits, but if the qubits are physically arrayed on a two-dimensional grid with $n$ qubits, then the distance between the data qubits and check qubit can scale as approximately $\sqrt{n}$. To implement this on a local architecture with static qubits, one needs either 2$\sqrt{n}$ swaps or additional qubits to generate teleportation channels that also require $\sqrt{n}$ operations \cite{Rosenbaum2012}.  Consequently the chance of a single check operator being  error free is exponentially smaller in the physical distance than that of the non-local architecture.

High-fidelity magic states, when combined with a teleportation circuit, can be consumed to generate a unitary operation on the logical data that is difficult to directly implement in the codespace  \cite{BravyiMagic}.   Distilllation techniques, which rely on one or more additional quantum error correction codes, provide a means for producing high-fidelity magic states from low-fidelity magic states \cite{Trout2015}.  This distillation process can be resource intensive for high-precision algorithms. Recently, sets of codes have been discovered that do not require magic states for universal quantum computation. Whether these codes present resource reductions relative to the surface code, depends on the code thresholds and the algorithm of interest.  The 3D color code is a promising code that does not require magic states  and but the QEC graph can be mapped to three dimensional space and not a surface .  We note that recently a dual 2D lattice code which in some sense is a stacked 2D color code, also has a simple set of universal gates \cite{Bravyi2015,Jochym2015}.

The modularity and reconfigurability of the ion trap and photonic architecture presented here is ideal for testing these and future codes.  Both surface codes \cite{Benjamin2013} and concatenated CSS codes \cite{MUSIQC} have already been mapped to the modular ion trap architecture considered here, with complementary features. Unlike architectures with static connections built for specific codes, any code can be implemented without constructing new hardware. It is worth noting that the surface code was invented only a few years after the first concatenated codes, but an additional decade  of research was required before it could be shown that the overhead was favorable compared to concatenated CSS codes for large algorithms. As hardware improves we expect that there will be continual improvements in quantum coding theory.

\section{Outlook}
Quantum computers will look very different than the semiconductor-based computers of today, just as current solid-state semiconductor devices look nothing like the vacuum tubes, relays, and mechanical gears of an earlier era.  While trapped atomic ion qubits may be seen as exotic today, their exquisite quantum coherence, high-performance quantum logic gates, and unmatched connectivity and reconfigurability makes the trapped ion platform a leading candidate for large-scale quantum computing.  The continued progress in ion trap integration strategies and supporting technologies have the potential to enable practical quantum computing machines in a matter of years.  We expect this device development to be driven by applications that harness the connectivity and reconfigurability of trapped ion qubits, where quantum computer scientists work closely with physicists and engineers in the co-design of tomorrow's quantum computer. 

\bibliography{NatureReview}

\section{Acknowledgements}
This work is supported by the U.S. Army Research Office (ARO) with funds from the IARPA MQCO Program and the ARO Atomic and Molecular Physics Program, the AFOSR MURI on Quantum Measurement and Verification,  the DARPA Quiness Program, the Army Research Laboratory Center for Distributed Quantum Information, the NSF Physics Frontier Center at JQI, and the NSF Physics at the Information Frontier program.

\end{document}